\documentclass[lettersize,journal]{IEEEtran}
\usepackage{amsmath,amsfonts}
\usepackage{algorithmic}
\usepackage{algorithm}
\usepackage{multicol}
\usepackage{array}
\usepackage[caption=false,font=normalsize,labelfont=sf,textfont=sf]{subfig}
\usepackage{textcomp}
\usepackage{stfloats}
\usepackage{url}
\usepackage{verbatim}
\usepackage{graphicx}
\usepackage{cite}
\usepackage[justification=centering]{caption}
\begin{document}

\title{ISAC Channel Modelling - Perspectives from ETSI}

\author{\IEEEauthorblockN{Mohammad Heggo, Arman Shojaeifard, Alain Mourad\\}
\IEEEauthorblockA{
\IEEEauthorblockN{\textit{InterDigital, Inc., London, UK}}\\
\vspace{0.5em}}
\and
\IEEEauthorblockN{Chuangxin Jiang, Ruiqi Liu, Junchen Liu\\}
\IEEEauthorblockN{\textit{ZTE Corporation, Shenzhen, China}\\
}
}
\maketitle
\markboth{ISAC Channel Modelling - Perspectives from ETSI}%
{ISAC Channel modelling - Perspectives from ETSI}


\begin{abstract}
Integrated Sensing and Communications (ISAC) is defined as one of six usage scenarios in the ITU-R International Mobile Telecommunications (IMT) 2030 framework for 6G. ISAC is envisioned to introduce the sensing capability into the cellular network, where sensing may be obtained using the cellular radio frequency (RF) signals with or without additional auxiliary sensors. To enable ISAC, specification bodies such as European Telecommunications Standards Institute (ETSI) and Third Generation Partnership Project (3GPP) have already started to look into detailed ISAC use cases, their requirements, and the channel models and evaluation methodologies that are necessary to design and evaluate ISAC performance. With focus on the channel model, the current communication-centric channel models like those specified in 3GPP technical report (TR) 38.901 do not cover the RF signals interactions between the transmitter, target object, receiver and their surrounding environment. To bridge this gap, 3GPP has been looking into the basic changes that are necessary to make to their TR38.901 channel model with focus on selected use cases from the 3GPP SA1 5G-Advanced feasibility study. In parallel, ETSI ISAC Industry Specification Group (ISG) has been studying the more advanced ISAC channel modelling features that are needed to support the variety of ISAC use cases envisioned in 6G.  In this paper, we present the baseline and advanced features developed thus far in 3GPP and ETSI ISAC ISG, respectively, towards a comprehensive view of the ISAC channel model in 6G.
\end{abstract}


\section{Introduction}
In November 2023, the International Telecommunication Union (ITU) released the International Mobile Telecommunications (IMT) framework IMT-2030 \cite{sector2023recommendation}, marking a pivotal step in the establishment of standards for the forthcoming sixth generation (6G) mobile communication systems. This framework delineates significant advancements over the existing fifth generation (5G) systems, specifically in areas such as data rate, throughput, sustainability, and the integration of artificial intelligence (AI) capabilities. A comprehensive overview of advancements in 6G technology is detailed in \cite{10904090}.

Moreover, the framework emphasizes enhanced sensing capabilities and introduces a range of innovative use case scenarios that necessitate integrated sensing and communication (ISAC) functionalities \cite{9737357,9945983}. These scenarios encompass assisted navigation, activity recognition, movement tracking, and environmental monitoring, among others. It is important to note that ISAC has advanced significantly and has yielded promising results in recent experimental trials \cite{10286534}. The IMT framework lays the groundwork for a transformative evolution in mobile communications, positioning 6G to meet the demands of future applications and technological ecosystems.

The standardization of ISAC has attracted significant interest from the academic community in recent years, focusing on various aspects such as key technological challenges, enabling factors, and resource allocation strategies \cite{10529727,10233791}. In 3GPP, a comprehensive feasibility study on ISAC has been carried out by the system architecture sub-group SA1. The group has published its latest report, TR 22.837 \cite{3gpp2024study}, which outlines 32 detailed use cases for ISAC. This report meticulously describes the service flows, preconditions, postconditions, functionality requirements, and key performance indicators (KPIs) associated with each use case. The ISAC use cases encompass a wide range of applications, including intruder detection, collision avoidance, automotive maneuvering and navigation, public safety, rainfall monitoring, and health monitoring. These use cases effectively leverage wireless sensing capabilities along with the functionality of non-3GPP sensors, such as cameras, microphones, and radars, thereby addressing a variety of real-world challenges.

The integration of ISAC features within contemporary mobile communication systems necessitates an in-depth comprehension of the ISAC channel dynamics. Currently, the latest version of the mobile communication channel model specified in TR 38.901 \cite{3gpp2024ISAC} lacks provisions for modelling the ISAC channel. To tackle this gap within the 3GPP technical specification group radio access network (TSG RAN), 3GPP RAN1 has initiated a study item aimed at exploring the ISAC channel model, with a focus on assessing the implementation of ISAC features. This investigative effort encompasses the modelling of various components of the ISAC channel, including: 1) diverse target object categories (such as humans, unmanned aerial vehicles (UAVs), and automated guided vehicles (AGVs)), 2) various sensing modes (like bistatic and monostatic modes), and 3) distinct modelling parameters tailored to specific use case scenarios.

In November 2023, the European Telecommunications Standards Institute (ETSI) established the ISAC Industry Specification Group (ISG) to explore the integration of ISAC features within 6G mobile communication systems. This initiative comprises five principal work items that collectively provide an in-depth analysis of ISAC implementation.

The first work item delineates the use cases for ISAC in the 6G context, identifying pertinent deployment scenarios and prospective frequency bands. It further articulates the requirements, key performance indicators, and value metrics associated with these use cases. This work item's findings culminated in a comprehensive group report published in February 2025, expanding the recognized ISAC use cases beyond the original 32 identified by 3GPP SA1. The second work item focuses on the development of advanced channel models for ISAC. This initiative complements ongoing research efforts in channel modelling as noted in 3GPP, while also drawing insights from established ISAC channel modelling initiatives within the academic community \cite{10333766,10592557}. The third work item investigates enhancements to the mobile system and RAN architecture to facilitate the integration of sensing capabilities. This work entails an analysis of various sensing types, integration levels, and deployment strategies tailored for the 6G system. The fourth work item concentrates on delineating security and privacy requirements pertinent to the identified 6G ISAC use cases, addressing potential vulnerabilities and safeguarding measures. Finally, the fifth work item explores critical issues related to ISAC computing within a 6G framework, examining the interplay among computing, sensing, and communication functions to optimize system performance and efficiency.

This paper summarizes the challenges of modelling the ISAC channel, including the gaps in 3GPP defined  mobile communication channel model in TR 38.901 and corresponding efforts by 3GPP to bridge these gaps by developing ISAC channel models. We provide an overview of the advanced ISAC channel modelling features developed in ETSI ISG ISAC, including relevant synergies with respect to the ongoing work in 3GPP. Conclusions and future directions are also provided towards the end.

\section{Gaps in 3GPP Communication Channel Models}
\subsection{Analyzing Deficiencies in Communication Channel Models}
The channel model referenced as 3GPP prior to Rel-19 incorporates both stochastic modeling and a hybrid approach that combines deterministic modeling (ray-tracing) with stochastic techniques. This model is intended for communication links from the base station to the user equipment (UE) and vice versa; however, it does not address the requirements for sensing channel modeling. It is parameterized for a variety of communication scenarios, but it lacks the consideration of various sensing scenarios, including monostatic and bistatic sensing channels, rendering it inadequate for future sensing applications.

The ISAC channel modeling requirements are typically contingent upon specific use cases (UCs) and associated performance criteria. Notably, several aspects necessitate updating to adequately accommodate sensing applications:
\begin{enumerate}
    \item \textit{Support for Monostatic and Bistatic Sensing:} Different sensing scenarios involve either co-located transmitter (Tx) and receiver (Rx) devices (monostatic) or separate locations for these components (bistatic). For monostatic sensing, both the Tx and Rx are within the same device, such as a base station (BS) or UE. In contrast, bistatic sensing may occur in configurations such as BS-to-UE, UE-to-BS, UE-to-UE, or BS-to-BS sensing modes, necessitating distinct channel modeling parameters for effective sensing channel generation.
    \item \textit{LoS/NLoS State Determination for Sensing Channels:} Conventional radar-based sensing channel models typically assume a line-of-sight (LoS) condition between the sensing device and the target. While this simplification may suffice for evaluating basic sensing performance, a comprehensive assessment of ISAC performance requires a model that includes both LoS and non-line-of-sight (NLoS) conditions. Existing distance-dependent LoS/NLoS decision probability models may be adapted to account for the spatial relationship between the Tx sensing device and the target, alongside the distance to the Rx sensing device.
    \item \textit{Large-Scale Parameters (LSPs) for Sensing Channels:} Variables such as delay spread (DS), angular spread (AS), shadowing factor (SF), path loss (PL), and Ricean K factor (K) may differ from those in communication channels. In communication scenarios, varying antenna heights between the base station and UE inform the use of different angular distributions for angle-of-arrival (AoA) and angle-of-departure (AoD) metrics. In monostatic sensing, however, since the Tx and Rx reside within the same device at the same height, the AoA and AoD can be modeled using a single distribution.
    \item \textit{Small-Scale Parameters (SSPs) for Sensing Channels:} This involves cluster or ray modeling tailored for sensing applications, taking into account the impact of mobility. Deterministic or semi-deterministic modeling may be employed; in the deterministic case, cluster locations are defined based on the known physical characteristics of the target or clutter. Conversely, semi-deterministic approaches utilize stochastic methods to characterize these objects. Channel metrics such as delays, gains, and echo angles for direct paths to sensing targets should be derived from physical locations, while indirect paths may be generated statistically based on empirical data.
    \item\textit{Target Modeling Considerations:} The radar cross-section (RCS) of relevant sensing targets must be thoroughly characterized. Different targets like humans, vehicles, or automated guided vehicles (AGVs) will exhibit varying RCS values. Additionally, micro-Doppler modeling or variable RCS strategies may be necessary for health monitoring signals, such as those associated with respiration and heartbeat. Advanced sensing applications that focus on target identification, posture recognition, and orientation detection require a geometric representation of targets, meaning multiple scattering centers with distinct RCS models should replace the simplistic single RCS value.
    \item \textit{Background Channel Considerations:} Understanding the interaction between the background environment and sensing targets is crucial for accurately evaluating sensing performance. Reflections from background objects, alongside those from primary targets, can contribute significantly to received power and consequently influence the overall sensing effectiveness. 
\end{enumerate}

In conclusion, advancing channel modeling to support sensing will necessitate a comprehensive reevaluation and enhancement of existing frameworks to address these emerging needs in integrated sensing and communication strategies.
\subsection{ISAC Channel Modelling Efforts in 3GPP}
The 3GPP RAN1 group launched a study item that aimed at developing a comprehensive channel model to support specific ISAC use cases and their associated requirements, as outlined in 3GPP SA1 TR 22.837 and TS 22.137. This study endeavors to establish a unified channel modeling framework that enables the detection and tracking of the specific targets (e.g. UAVs, humans, vehicles, AGVs), while also differentiating them from unintended targets (e.g. hazards, environmental objects).

The scope of the study encompasses all six sensing modes: BS-BS bistatic, BS monostatic, BS-UE bistatic, UE-BS bistatic, UE-UE bistatic, and UE monostatic. The initial focus will be on the frequency range of 0.5 – 52.6 GHz, with the expectation that the model will be extendable to 100 GHz. The study prioritizes enhancements to the stochastic model presented in 3GPP TR 38.901, particularly with regard to incorporating known characteristics of physical objects, in contrast to alternative modeling approaches such as map-based hybrid channels or ray-tracing techniques.

The proposed channel modeling framework includes the following components:
\begin{itemize}
    \item The ISAC channel comprises both a target channel and a background channel.
    \item  Multiple sensing targets can be modeled within the ISAC channel of a given transmitter-receiver (Tx-Rx) pair. Also, a single sensing target may be represented across the ISAC channels of various Tx-Rx pairs.
    \item  NLOS components within the target channel can be modeled using stochastic clusters or environmental objects (EOs), classified as follows:
    \subitem EO type-1, which follows a similar model to that of a sensing target.
    \subitem EO type-2 (e.g., ground, wall, ceiling), which is represented through specular reflection.
\end{itemize}

For the target channel, consensus has been reached on the following points:
\begin{itemize}
    \item For each Tx-target and target-Rx link, the state may be either Line of Sight (LOS) or NLOS. The channel model parameters for each link can be represented based on legacy communication channel model in TR 38.901.
    \item These states are generated by concatenating the parameters of the Tx-target link with those of the target-Rx link. Existing channel model presented in TR 38.901 can be utilized for both links.
\end{itemize}

Modelling approaches for physical objects may leverage either single point or multiple scattering point techniques, although the relationships concerning use cases, object dimensions, target distance, and other variables are still under deliberation. Notably, progress has been made regarding the modeling of the object's RCS, with an agreement on a scalar RCS value combined with a complex-valued polarization matrix (CPM). Nonetheless, discussions around the specific definition of the RCS and CPM are ongoing.

Currently, the 3GPP RAN1 study has yet to achieve consensus on a comprehensive model that encompasses two critical attributes of the target: 1) RCS and 2) Micro-Doppler(MD). These elements are pivotal for the accurate detection and identification of targets, as they are distinguishing features of sensing targets that can be used to improve detection and tracking performance in the identified use cases. As previously mentioned, the study has deprioritized the use of ray-tracing and various simulation tools, including electromagnetic tools, in assessing the performance of the ISAC channel. While these tools may introduce additional complexity to the current communication channel model, they possess the capability to achieve a more precise representation of the target parameters and the interactions between the target and its surrounding environment. This situation creates opportunities for collaborative initiatives by other pre-standardization bodies, such as ETSI, to address these identified gaps. 
\section{ISAC Channel Model Features from ETSI}
\subsection{RCS Modelling}
The RCS is defined as the equivalent area of a perfectly reflecting sphere that would produce the same radar reflection strength as the target object. It quantifies a target's ability to backscatter radar signals towards the receiver and is integral in assessing detectability based on electromagnetic wave incidence on the object. Accurate RCS measurement and modeling are essential for characterizing the ISAC channel, as different sensing target categories (e.g. human, vehicles, etc.) respond uniquely to electromagnetic fields. This subsection outlines the key requirements for RCS modeling in the ISAC context and presents ETSI ISAC ISG framework for modeling various target objects in the environment.

The RCS of a target is a defining characteristic influenced by several parameters:
\begin{itemize}
    \item Frequency of the incident radio signal,
    \item Angles of incidence and scattering: 
    \begin{itemize}
        \item Angle of incidence: the angle between the line of sight (LoS) path from the transmitter to the target and the target surface normal,
        \item Scattering angle: the angle between the LoS path from the receiver to the target and the target surface normal.
    \end{itemize} 
    \item Polarization of both transmitter and receiver,
    \item Characteristics of the target object, such as: material, size, shape, motion or sub-motion, and orientation.
    \item Antenna pattern of Tx/Rx,
    \item Distance from either Tx or Rx.
\end{itemize}
These factors collectively impact the RCS modeling of the target.

The modeling of object RCS can be categorized into two primary approaches: top-down and bottom-up. In the top-down approach, the RCS model is established by measuring the ratio of scattered to incident electric fields from the object. These measurements can be sourced from experimental campaigns, numerical simulations, or ray-tracing methods. Conversely, the bottom-up approach derives closed-form expressions for scattered and incident electric fields utilizing electromagnetic propagation theory. For practical applicability, this approach often requires calibration against empirical data or numerical simulations. Top-down methods are particularly effective for complex objects, such as humans and vehicles; however, their main limitation is the narrow applicability of the resulting RCS models to specific scenarios. In contrast, bottom-up approaches excel in modeling fundamental shapes—rectangles, spheres, dihedrals—offering scalability regarding size, frequency, and material properties. Nevertheless, they struggle with the extension to complex geometries.

The ETSI ISAC ISG is employing a novel multipoint representation approach for modeling RCS, emphasizing target object segmentation as a primary technique. This method involves breaking down complex objects into simpler components with known RCS values to enhance modeling accuracy. Examples from previous literature include segmenting a traffic sign into a cylinder and a square for its pillar and sign, as well as disassembling wind turbine blades into rectangular segments to evaluate their RCS \cite{6573336}\cite{7236922}. The segmentation approach accounts for multiple reflection points and enables the calculation of RCS for electrically large objects, which can be refined through empirical data or advanced numerical simulations like computer simulation technology (CST) studio suite software. This strategy allows for the development of RCS models adaptable to various frequencies and object sizes, facilitating the assessment of RCS at differing transmitter/receiver distances. 

In the segmentation methodology, a target object is divided into $N$ segments, each characterized by an RCS value, denoted as $\sigma_{i}$, which can be determined based on the target segment's parameters such as geometric configuration (such as circular or rectangular), dimensions, material and operational frequency. The average RCS of the object $\sigma_{obj}$ can be derived as a function of the RCS of the individual segments $\sigma_{i}$. The computation of the RCS for the target object relies on whether the incident or scattered electromagnetic wave can be categorized as a plane wave. The validity of the plane-wave assumption is contingent upon the interplay of three critical parameters: 1) the maximum dimension of the object $D^{obj}_{max}$, 2) the wavelength of the incident radio signal $\lambda$, and 3) the minimum distance between the target and the transmitter $r_{tx}$ or receiver $r_{rx}$ referred to as $\min(r_{tx},r_{rx})$. For the scattered wave to be considered a plane wave, the minimum distance $\min(r_{tx},r_{rx})$ must exceed the ratio of $\frac{2{D^{obj}_{max}}^2}{\lambda}$. Should this criterion be unmet, the plane wave approximation becomes invalid. In such scenarios, the segmentation approach is employed, which entails dividing the scattering object into $N$ segments. The minimum quantity of segments must adhere to the plane wave assumption, while the maximum segments number must ensure that the object's maximum dimension surpasses the wavelength of the incident signal.

\begin{figure}[htbp]
    \centering
    \includegraphics[width=0.45\textwidth]{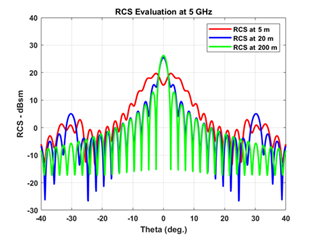}
    \captionsetup{justification=centering}
    \caption{RCS at different aspect angles and separation distances from the sensing receiver}
    \label{fig:RCS}
\end{figure}

The RCS of a metallic plate was analyzed using a segmentation approach across various orientations and separation distances from the sensing receiver as shown in Fig. \ref{fig:RCS}. According to the simulation results in the ETSI ISAC ISG early version report, the target RCS can be primarily categorized into two components: 1) Slow fading and 2) Fast fading. The slow fading component is influenced by target-specific parameters, such as shape, size, material, and distance from the Tx/Rx, in addition to incident signal characteristics like frequency and polarization. Conversely, the fast fading component is determined by the incidence/scattering angle relative to the sensing target, which changes with the target's orientation toward the Tx/Rx. This component effectively models the RCS directivity gain in specific directions.

To integrate RCS within the existing communication channel framework outlined in TR 38.901 (e.g. the system level simulator), the channel large-scale parameter (LSP) and small scale parameter (SSP) procedures have been modified to include both slow-fading and fast-fading components of RCS, as depicted in Figure \ref{fig:RCS}. The initial adjustment pertains to the path loss (PL) model, which has been enhanced to accommodate the slow-fading component of RCS. The subsequent modification involves the evaluation of cluster angles, which has been expanded to incorporate the fast-fading RCS component. Finally, the channel coefficient generation block has been updated to integrate the RCS fast-fading component within equations 7.5-28 and 7.5-29 in TR 38.901.

\subsection{Micro-Doppler Modelling}
In the context of ISAC channels, the micro-Doppler effect refers to the frequency or phase alterations in signals reflected from a target, caused by subtle movements such as rotation, vibration, or hand gestures. This effect results in sideband shifts around the primary Doppler frequency shift due to the target's bulk motion and varies over time \cite{chen2019micro}. Key factors influencing the micro-Doppler effect include:
\begin{enumerate}
    \item \textit{Amplitude of Fine Motion:} Depends on the extent of motion (e.g., vibration) or target geometry (e.g., propeller size).
    \item \textit{Frequency of Fine Motion:} Can consist of one or multiple frequency components, each with distinct amplitudes.
    \item \textit{Periodicity of Motion:} Classifies motion as periodic (e.g., walking, rotation) or aperiodic (e.g., hand gestures).
    \item \textit{Carrier Frequency:} The ratio of fine motion amplitude to carrier wavelength affects the micro-Doppler signature; shorter wavelengths yield greater frequency shift amplitudes.
    \item \textit{Motion Direction:} The angle between the incident signal and the target's movement influences the micro-Doppler effect.
\end{enumerate}
This compact delineation highlights the complex interactions of micro-Doppler parameters relevant to sensing applications.

In TR 38.901, the mobile communications channel model defines the Doppler motion associated with either the transmitter or receiver of the mobile RF signal. Within the 3GPP RAN1 ISAC study item, a consensus was reached to characterize the Doppler motion of the target in conjunction with its micro-Doppler motion. However, this study item lacks a comprehensive model for the micro-Doppler motion of the target. The early version group report from ETSI ISAC ISG for second work item  proposes an accurate micro-Doppler motion model that integrates both the Doppler and micro-Doppler motions of the target. This model is designed for integration with the mobile communication channel model outlined in TR 38.901. Furthermore, the ISG has introduced various micro-Doppler models that effectively capture distinct micro-Doppler phenomena, including helicopter propeller rotation, UAV propeller dynamics, vibrational motion, as well as human activities such as walking, breathing, and heartbeat patterns.

To validate the MD model and establish its compatibility with TR 38.901, simulations were conducted for the sensing of micro-motion utilizing the link-level system channel model specified in TR 38.901. In this simulation, the ISAC channel is partitioned into sensing clusters, each characterized by varying normalized delays, incident power levels, and angles of arrival and departure of corresponding RF rays. A single cluster is modeled as a sensing cluster related to a target moving at a speed of 150 km/h, while the remaining clusters are treated as environmental clusters with zero velocity. In addition to the macro-motion, a random subset of rays within the sensing cluster is assigned the same micro-Doppler phase in accordance with the defined micro-Doppler function. Two distinct micro-motions are explored within the simulation, each represented by a different micro-Doppler function. Detailed parameters of the simulation are outlined in Table \ref{table:table}.

\begin{table}
\caption{Simulation Parameters}
\label{table:table}
\begin{center}
\begin{tabular}{|c|c|}
\hline
\textbf{Parameter} & \textbf{Value} \\
\hline
Carrier frequency & 3.5 GHz \\
\hline
Sub carrier space & 30 kHz \\
\hline
Bandwidth of sensing signal & 18 MHz \\
\hline
Number of rays in one cluster & 20 \\
\hline
Pulse repetition interval of sensing signal &	14 OFDM symbol\\
\hline
Time length of sensing signal &	1 OFDM symbol\\
\hline
Sensing signal repetition number &	50\\
\hline
Signal to noise ratio (SNR) &	30dB\\
\hline
Speed of sensing target	& 150km/h\\
\hline
Micro-motion mode 1 speed & 50\\
\hline
Micro-motion mode 1 shape & cosine\\
\hline
Micro-motion mode 2 speed & 30\\
\hline
Micro-motion mode 2 shape & sawtooth\\
\hline

\end{tabular}
\end{center}
\end{table}

\begin{figure}[htbp]
    \centering
    \includegraphics[width=0.45\textwidth]{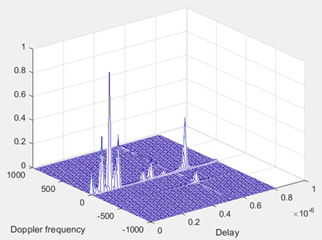}
    \captionsetup{justification=centering}
    \caption{Delay-Doppler profile of ISAC channel}
    \label{fig:delay_doppler1}
\end{figure}
The estimated channel state information (CSI) from the received target echo signal is presented in Fig. \ref{fig:delay_doppler1}. Analysis reveals that the environmental channel exhibits significant power near the zero Doppler frequency, which can be effectively mitigated using a zero notch filter. By suppressing the environmental channel at the zero Doppler point, we can preserve the integrity of the sensing channel associated with the target, which operates at higher Doppler frequencies.

A sequence of Doppler vector data related to the target is consolidated into a delay-Doppler profile for target detection. Analysis of the Doppler values across two distinct micro-Doppler modes reveals that these modes exhibit unique Doppler patterns, which can be leveraged for classification purposes. 
\section{Evaluation Framework for the ISAC Channel model}
Evaluation methodologies are expected to be studied in both 3GPP and ETSI following the development of ISAC channel models. Here, we provide initial findings from ETSI related to conducting measurement campaigns and utilizing ray-tracing (RT) tools for human MD motions.
\subsection{Measurement Results}
\begin{figure}[htbp]
    \centering
    \includegraphics[width=0.45\textwidth]{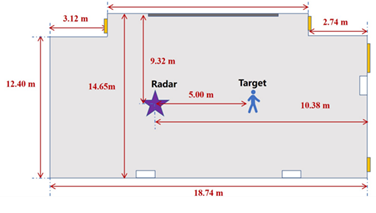}
    \captionsetup{justification=centering}
    \caption{Dimensions of the room and radar position}
    \label{fig:MeasurementSetup}
\end{figure}

To minimize external clutter, a pristine indoor environment was chosen as the experimental setting for this study. As illustrated in Fig. \ref{fig:MeasurementSetup}, the room has a height of 2.73 m and an approximate total area of 250 m². The radar apparatus is centrally positioned within the space and directed towards one of the walls. Prior calculations indicated that the linear velocity of the participant's swinging arm is approximately 2 m/s. The study accounted for a participant height of 1.75 m, with the individual standing directly in front of the radar and solely executing arm-swinging motions. This approach mitigates the effects of echoes resulting from bodily movements, thereby allowing for a clearer observation of the Doppler characteristics associated with micro-motion. The participant mimics a natural walking gait pattern, positioning the left arm posteriorly while advancing the right arm. This motion is reciprocated, with the right arm transitioning to the back and the left arm moving forward, creating a circular motion. A total of three arm-swing cycles, corresponding to a duration of three seconds, were recorded for analysis.

\begin{figure}[htbp]
    \centering
    \includegraphics[width=0.45\textwidth]{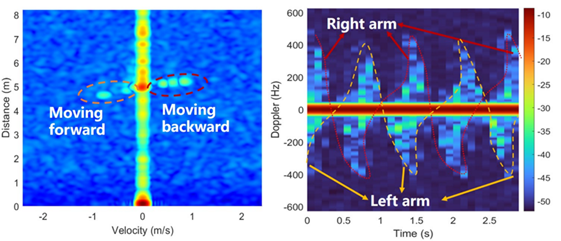}
    \captionsetup{justification=centering}
    \caption{Measurement result: Velocity-Distance plot (left) and delay-Doppler plot (right)}
    \label{fig:MeasurementResults}
\end{figure}

The results are illustrated in Fig. \ref{fig:MeasurementResults}. The velocity-distance figure on the left indicates negative velocity for the arm advancing towards the radar, while positive velocity is recorded for the arm retreating from it. On the right side of Fig. \ref{fig:MeasurementResults}, the delay-Doppler figure reveals a noticeable periodicity in the movements of both the left and right arms. Furthermore, the oscillating arm demonstrates a remarkably consistent pattern over continuous time.
\subsection{RT Simulation Results}
\begin{figure}[htbp]
    \centering
    \includegraphics[width=0.45\textwidth]{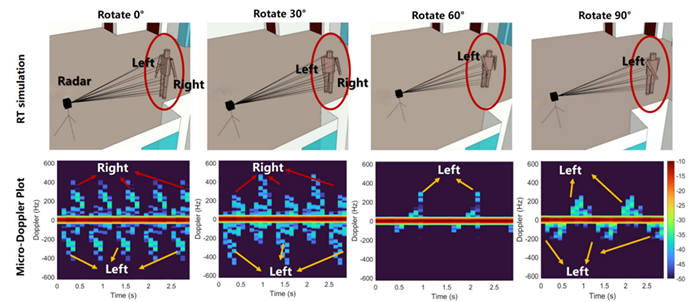}
    \captionsetup{justification=centering}
    \caption{RT simulation results of human motion at different orientation angles}
    \label{fig:RTResults2}
\end{figure}

The RT simulation tool has been utilized to model the dynamics of human arm motion during swinging activities (e.g. human walking). This simulation framework provides enhanced flexibility for exploring a variety of use case scenarios, including different target orientations relative to the RF radar system. For instance, as illustrated in Fig. \ref{fig:RTResults2}, the results from the RT simulation reveal that the micro-Doppler signature of the swinging arms varies with the rotational angle of the body. Notably, the regularity of these variations depends on both the radial distance from the radar and the speed of motion. When the trajectory of the hands aligns parallel to the radar's line of sight, the resulting Doppler pattern is pronounced, exhibiting a significant amplitude. Conversely, when the hands are moved perpendicularly to the radar's line of sight, such as in a rotated position, the detected Doppler shift is minimal. This phenomenon produces a tooth-like waveform in the signal, attributed to the acceleration and deceleration phases during the arm waving motion.

\section{Insights and Future Directions}
This paper provided an overview of the ISAC channel modelling activities in ETSI, with a focus on proposed methodologies for RCS and MD modeling, and relevance to ongoing work in 3GPP. 

A segmentation approach has been proposed as a method for constructing an RCS model for complex objects based on the known RCS of fundamental objects. This model can be scaled to accommodate various frequency bands and object dimensions. Furthermore, segmentation allows for the evaluation of the RCS of electrically large objects as a function of varying separation distances from the transmitter/receiver. Future iterations of the GR may explore additional methodologies for RCS modeling that incorporate electromagnetic approaches. While these methods can offer greater accuracy, they may necessitate enhanced computational capabilities. Electromagnetic simulation tools can also yield precise RCS models for more intricate targets, such as humans and drones. 

Moreover, detailed micro-Doppler modeling is proposed based on empirical measurement campaigns and real-time emulations. The patterns generated through this process can inform use cases in human motion recognition, real-time health hazard monitoring, disaster risk assessment, emergency search and rescue operations, as well as healthcare sensing and monitoring. Future work may extend to modeling micro-scale variations in targets, such as micro-deformations in structural elements like bridges or buildings. Additionally, the study of ISAC channel modeling may encompass advanced topics, including the near-field effects, particularly at higher frequency bands (e.g., FR2 or FR3).

Following the latest progress on developing ISAC channel models in ETSI and 3GPP, validation efforts through comprehensive measurement campaigns and emulations are needed. Additionally, frameworks for evaluation methodology and  feasibility analysis are required to assess the performance of ISAC systems. These initiatives must be explored within the broad ecosystem including standardization bodies, individual industry members, stakeholder associations, academia, and strategic collaborative projects at national and regional levels.
\section{Conclusion}
In this paper, we provided an overview of research and standardization activities from ETSI around ISAC channel modelling, with a particular emphasis on the RCS and MD modeling approaches. Both models underwent rigorous validation through extensive measurement and simulation campaigns. The RT simulation results demonstrate their effectiveness in capturing complex scenarios within a dynamic ISAC channel environment. Furthermore, we discussed the future directions for ISAC channel modeling in view of the latest progress in ETSI and 3GPP, highlighting the need for a deeper investigation into measurement campagins and emulations. It is  recommended to integrate the proposed ISAC channel models into forthcoming research and specification endeavors pertaining to 6G ISAC systems.



\end{document}